\documentstyle[prl,aps,twocolumn]{revtex}

\begin{document}

\title{Correlated-Electron Theory of Strongly Anisotropic Metamagnets}

\author{K. Held$^{a}$, M. Ulmke$^{b,}$\thanks{Present
address: Physics Department, University of California, Davis, CA 95616, USA},
 and D.~Vollhardt$^{a}$}

\address{$^{a}$Institut f\"{u}r Theoretische Physik C,
RWTH Aachen, D-52056 Aachen, Germany \\
$^{b}$Institut f\"{u}r Festk\"{o}rperforschung, Forschungszentrum J\"{u}lich,
D-52425 J\"{u}lich, Germany}

\date{May 23, 1995}
\maketitle

\begin{abstract}
We present the first correlated-electron theory
of metamagnetism in strongly anisotropic antiferromagnets.
 Quantum-Monte-Carlo techniques are used
to calculate the field vs. temperature phase
diagram of the infinite-dimensional Hubbard model
with easy axis.
A metamagnetic transition scenario with 1.~order and 2.~order phase
transitions is found. The apparent
similarities to the phase diagram of FeBr$_2$ and
to mean-field results for the Ising model with competing interactions
are discussed.

\end{abstract}


In 1939 Becquerel and van den Handel \cite{1} reported
the observation of an anomalous magnetization behavior in the mineral
mesitite (carbonate of Fe and Mg) at low temperatures which they could not
explain in terms of existing theories of magnetism. Therefore they
suggested for it the name ``metamagnetism'' \cite{2}. Qualitatively
similar, but  even more drastic magnetization effects
 were later observed in many other systems of which
FeCl$_2$ and Dy$_3$Al$_5$O$_{12}$ (DAG) are well-studied
prototypes \cite{3}. These materials are strongly
anisotropic antiferromagnets where the spins are constrained to lie
along an easy axis $\vec{e}$. Under the influence of a strong magnetic
field along $\vec{e}$ they undergo a 1.~order (``metamagnetic''),
rather than a spin-flop, transition
from the antiferromagnetic (AF) to a paramagnetic
 state. Today the term ``metamagnetic transition'' is used
 in a much wider sense \cite{4}, namely whenever $\chi (H)$ has a maximum
 at some value $H_c$,
with $M (H)$ being strongly enhanced for $H > H_c$ \cite{5}.
Metamagnetism is then found to be a rather common
phenomenon which occurs also in
strongly exchange-enhanced paramagnets
(e.g. \cite{4} TiBe$_2$, YCo$_2$), heavy fermion and intermediate valence
systems (e.g. CeRu$_2$Si$_2$, UPt$_3$), and \cite{5} the
parent compound of high-T$_c$ superconductivity, La$_2$CuO$_4$.

In this paper we will be concerned only with
metamagnetism in strongly anisotropic antiferromagnets, where a
spin-flop transition does not occur. Apart from the well-known insulating
materials, e.g. FeCl$_2$,  there are also some good
electrical conductors, especially UA$_{1-x}$B$_x$
(where $A =P$, $As$; $B = S$, $Se$), in this group \cite{3}.
Theoretical investigations of metamagnetism in these systems were sofar
restricted to the insulating systems, several of which
are known to have a very interesting $H-T$ phase diagram
($H$:  {\it internal} magnetic field, $T$: temperature).
In particular, it includes a tricritical point (TCP) at which
the 1.~order phase transition becomes 2.~order (Fig.~1a) -- a
feature very similar to that found in $^3$He-$^4$He mixtures \cite{6,7}.
Theoretical investigations of the multicritical behavior
 are usually based on the
Ising model with competing interactions \cite{7},  e.g.
\begin{equation}
H = J \sum_{NN} S_iS_j - J' \sum_{NNN} S_i S_j - H \sum_i S_i \; .
\label{Gl1}
\end{equation}
For $J, J' > 0$ one has  an AF coupling between the $Z$
nearest-neighbor (NN) spins and a ferromagnetic
coupling between the $Z'$ next-nearest-neighbors (NNN).
 It was  pointed out by
Kincaid and Cohen \cite{8}
that in mean-field theory a TCP as in Fig.~1a exists only for
$R \equiv Z'J'/(ZJ) > 3/5$, while for $R < 3/5$ this point separates
into  a critical endpoint (CE) and a bicritical endpoint (BCE)
(see Fig.~1b).
The latter behavior, especially the finite angle between the
two transition lines at CE and the pronounced maximum at the
2.~order line,
 is qualitatively very similar to that observed in FeBr$_2$ \cite{9}.
However, the 1.~order line between CE and BCE
has so far not been observed -- neither experimentally,
nor even theoretically when evaluating (\ref{Gl1})
beyond mean-field theory \cite{10,11}. Most recently, by
measurement of the excess magnetization and anomalous susceptibility
loss, de Azevedo et al. \cite{12} identified a strip-shaped regime
of strong {\it non-critical}  fluctuations in FeBr$_2$ which is at
least reminiscent of the {\it critical}
line CE $\leftrightarrow$ BCE.
This unusual behavior was then also found theoretically by
Selke and Dasgupta \cite{13} who evaluated (\ref{Gl1})
in the limit of  weak ferromagnetic coupling
$(J'/J = 0.2$) and large $Z$ using Monte-Carlo techniques.

It is the purpose of this paper to go beyond
 an effective, classical Ising-spin model and to investigate the origin of
metamagnetism in strongly anisotropic
antiferromagnets  from an even more microscopic point of view. Our
question is: which is the {\it minimal electronic
correlation model}
for metamagnetism in these systems? Since the
Hubbard model at half filling ($n = 1)$ is the generic microscopic model
for antiferromagnetism (both of itinerant and localized nature)
in correlated electron systems,
 we will study this model. For nearest-neighbor hopping of
electrons on a bipartite lattice in the presence of a magnetic field
it has the form
\begin{eqnarray}
\hat{H} & = & -t \sum_{NN, \sigma} \hat{c}_{i \sigma}^+ \hat{c}_{j \sigma}
 + U \sum_i \hat{n}_{i \uparrow} \hat{n}_{i \downarrow}  \nonumber \\
 & & - \sum_{i \sigma} (\mu + \sigma H) \hat{n}_{i \sigma}  \; ,
\label{Gl2}
\end{eqnarray}
where operators carry a hat. In this paper we consider a half-filled
band in which case $\mu = U/2$ by
particle-hole symmetry.
 From the exact, analytic solution in  dimensions $d = 1$
the (paramagnetic) ground state of this
model is known to exhibit metamagnetic behavior, i.e.
$\partial \chi/\partial H > 0$ up to saturation \cite{14}. The only
other dimension in which the dynamics of the correlation problem can be
treated exactly in the thermodynamic limit  is
$d = \infty$ \cite{15}. In this limit, with the scaling $t = t^*/\sqrt{Z}$
in (\ref{Gl2}), one obtains a dynamical
single-site problem \cite{16} which is
equivalent to an Anderson impurity model
complemented by a self-consistency condition \cite{17}, and is thus
amenable to numerical investigations within a finite-temperature
Monte-Carlo approach \cite{18}. This approach has recently provided valuable
insight into the physics of strongly correlated electron systems, e.g. the
Mott-Hubbard transition \cite{19} and transport properties \cite{20}. The
effect of the magnetic field $H$ in (\ref{Gl2}) for $n = 1$ was
also studied  \cite{21,22,23}. In particular, Laloux et al.
\cite{23} thoroughly investigated the magnetization behavior of the
paramagnetic phase, assuming the AF order to  be
suppressed. For $U = 3 \sqrt{2} t^*$  they find a 1.~order
metamagnetic transition between the strongly correlated metal
and the Mott-insulator at a critical field $H \simeq 0.2 t^*$.
Giesekus and Brandt \cite{21} also took into
account the AF order. They considered the
case where the field $\vec{H}$ orients
the staggered magnetization $\vec{m}_{st}$ perpendicular to itself.
For  isotropic antiferromagnets this is indeed the arrangement
with lowest energy. There is then no metamagnetism.

To investigate how metamagnetism may arise in correlated electron
systems with  strongly {\it anisotropic} AF order
we will work in the AF phase of the Hubbard model, too,
but will constrain $\vec{m}_{st}$ to lie {\it parallel} to $\vec{H}$.
In this way the existence of an easy axis $\vec{e}$ along which
$\vec{H}$ is directed, such that $\vec{e} \parallel
\vec{m}_{st} \parallel \vec{H}$, is incorporated
in a natural way. We will first discuss (\ref{Gl2}) in the limit
of strong and weak coupling since this will already show that the
appearance of a tri- or multicritical point is a delicate matter.

\underline{1. Strong coupling:} In the
limit $U \gg t$ the Hubbard model at $n = 1$
is equivalent to an effective Heisenberg spin-model,
$\hat{H}_{Heis} = J \sum_{NN} \sum_{\alpha = x,y,z} \hat{S}_i^\alpha
\hat{S}_j^\alpha - H \sum_i \hat{S}_i^z$, with an AF
exchange coupling $J = 4t^2/U$. For this model
Weiss-mean-field theory becomes exact
in $d = \infty$ yielding the same results as for the Ising model
(\ref{Gl1}). In this case the
transition line in the $H-T$ phase diagram has indeed the form shown in
Fig.~1a -- {\it but it is of 2. order for all} $T > 0$, i.e.
$T_t = 0$.  The
behavior changes if we
include the next term in the effective spin-Hamiltonian, of
order $t^4/U^3$, involving two- and four-spin terms \cite{24}.~It
leads to an effective  ferromagnetic spin
coupling $J'$.~Then we obtain
precisely the Ising model (\ref{Gl1}) in
mean-field theory with $J = 4t^2/U$ and $J' \propto
(t/U)^2 J \ll J$, and thus expect to find the transition scenario in
Fig.~1b. However, since the  expansion is in $t/U \ll 1$
 it is not clear down to what
values of $U/t$ this behavior can actually be observed.~

\underline{2. Weak coupling}: For $U \ll t$ we expect the Hartree-Fock
approximation, a {\it static} mean-field theory,
to give an, at least, qualitatively
correct answer.
Within this approximation we
find a metamagnetic phase transition, too --
{\it but it is of 1.~order for all  $T$ and} $U$,
corresponding to Fig.~1a, but
with $T_t = T_N$. Hence the Hartree-Fock solution cannot describe
the experimental situation. Apparently the location of the
(tri-) critical point in the $H$ - $T$ phase diagram,  and even
the transition scenario itself, depends sensitively on the value of
the electronic on-site interaction $U$. To study this point in
greater detail we have to go to intermediate coupling.

\underline{3. Intermediate coupling:} In this interaction
range we solve (\ref{Gl2}) numerically
in $d = \infty$. In contrast to Hartree-Fock this limit
provides a {\it dynamic} mean-field theory. The
 local self-energy $\Sigma_{\alpha n}^\sigma$ and propagator
$G_{\alpha n}^\sigma$, where the subscript $n$ denotes the
Matsubara frequency $\omega_n = (2n+1) \pi T$ and $\alpha \in
(A,B) = (+,-)$ is the sublattice index, are determined by
 two sets of dynamically coupled, self-consistent
equations for $G$ and $\Sigma$ \cite{16,17,25}:
\begin{equation}
G_{\alpha n}^{\sigma} = \int_{-\infty}^\infty d \epsilon
\frac{N^0 (\epsilon)}
{z_{\alpha n}^{\sigma} - \epsilon^2/z_{-\alpha n}^\sigma}
= - \langle \psi_{\sigma n} \psi_{\sigma n}^*
\rangle_{A_\alpha} \; .
\label{Gl4}
\end{equation}
Here $z_{\alpha n}^\sigma = i \omega_n +  \mu - \Sigma_{\alpha n}^\sigma$,
and the thermal average in (\ref{Gl4}) is defined as a functional integral
over the Grassmann variables $\psi, \psi^*$, with
$\langle {\cal O} \rangle_{A_\alpha} =
\int {\cal D} [\psi ] {\cal D} [\psi^*] {\cal O}
\exp ({A_{\alpha}\{ \psi, \psi^*,\Sigma, G\} }) /Z_\alpha$,
in terms of the
single-site action
\begin{eqnarray}
A_\alpha & = & \sum_{\sigma, n} \psi_{\sigma n}^*
[ (G_{\alpha n}^\sigma)^{-1} +
\Sigma_{\alpha n}^{\sigma} ] \psi_{\sigma n} \nonumber \\
 & & - U \int_0^\beta d \tau \psi_\uparrow^* (\tau) \psi_\uparrow (\tau)
 \psi_\downarrow^* (\tau) \psi_\downarrow (\tau) \; ,
\label{Gl5}
\end{eqnarray}
with $Z_\alpha$ as the partition function.
Furthermore
 $N^0(\epsilon)$ is the density of states of the non-interacting
electrons. As the results do not much depend on its precise form
we choose $N^0(\epsilon) = [(2t^*)^2 - \epsilon^2 ]^{1/2}/(2 \pi t^{*2})$.
 From now on we will set $2t^* \equiv 1$, i.e.
measure all energies in units of half the band width.
For finite $T$ and not too  large $U$ the
 functional integral
can be calculated numerically by Quantum-Monte-Carlo simulations
\cite{18,25}. Equ.(\ref{Gl4}) is solved by iteration.
All calculations were performed for $U= 2$ where the N\'{e}el temperature in
zero field  is close to its maximum value $T_{N, max} = 0.10$.

The results for  the magnetization $m(H)$ and the
staggered magnetization  $m_{st} (H)$ \cite{26}
are shown in Fig.~2. The metamagnetic behavior
is clearly seen: at
$T = 1/12 \; m(H)$ begins to show a typical
``S-shape'' which becomes more
pronounced at $T = 1/16$ (Fig.~2a).
For these temperatures
$m_{st}$ is a continuous function of $H$ (Fig.~2b), which vanishes at the
critical field $H_c$ as $m_{st} \sim (H_c - H)^\nu$. A least-square
fit through five points (weighted with the error bars) near the transition
yields $\nu = 0.46 \pm 0.05$ for $T = 1/12$, which is compatible
with a mean-field exponent $\nu = 1/2$.
By contrast, at the lowest temperatures, $T \leq 1/32$
 (Fig.~2c,d) the transitions
in $m$ and $m_{st}$ are clearly discontinuous.
Hysteresis is only weak  and is not shown. The field dependence
at intermediate temperatures $1/16 < T < 1/32$ is more complex:
 $m_{st}$ is almost field-independent (i.~e. $m \simeq 0)$
below some field $H'_c$,
decreases sharply at $H_c^\prime$ (such that $m > 0$), but
vanishes only at a field $H_c > H_c^\prime$
in a  continuous way.
Although the error bars do not permit an unambiguous interpretation
it seems that the order parameter decreases by {\it two} consecutive
transitions: the first one, at $H_c^\prime$, being of 1.~order and the second
one, at $H_c$, of 2.~order.

The results for $m(H)$ and $m_{st}(H)$
 are used to construct the $H-T$ phase diagram
shown in Fig.~3. It displays all the features of Fig.~1b.
In particular, the 1.~order line, defined by $H_c^\prime(T)$
 for $T > T_{CE}$,
 continues {\it into}
the ordered phase, separating two different AF phases: AF$_I$9412107
(where $ m \simeq 0$) and AF$_{II}$ (where $m > 0$).  The position
of its  endpoint  cannot, at
present, be determined accurately (dotted line).
 We note that the ratio of $T_{CE}$
and the N\'{e}el temperature $T_N$ at $H = 0$ is $T_{CE}/T_N
\simeq  0.3$.

The $H-T$ phase diagram in Fig.~3 is
strikingly similar to the experimental phase diagram of FeBr$_2$
\cite{12,27}.
In particular, the part of the 1.~order line extending into the ordered
phase clearly resembles the regime where de Azevedo et al.
\cite{12} observed an anomalous,
but {\it non}-critical, behavior in $m(T)$ and
$\chi (T)$. It appears
that  there may well exist a truely {\it critical}
feature in this part of the phase diagram. It would therefore be
interesting to measure $m_{st} (T,H)$ in FeBr$_2$
by neutron scattering to see whether
such a feature really exists.~--~The fact that
the results shown in Figs.~2,3 are qualitatively  very similar to
the mean-field solution of the Ising model (\ref{Gl1})
with $J^\prime \ll J$ \cite{8,11}
is surprising  since  at $U = 2$ (= bandwidth) the itinerant nature of the
electrons cannot yet be ignored. On the other hand,
the applicability of mean-field theory
itself to FeBr$_2$ seems
to be justified by the fact that in this
material the AF superexchange involves 20
equivalent sites in the two neighboring iron planes \cite{28}.
 We note that in FeBr$_2$
$T_{CE}/T_N = 0.33$~\cite{9}.

In summary,
we investigated the origin of metamagnetism in strongly
anisotropic antiferromagnets starting from
 a microscopic model of strongly
correlated-electrons, the Hubbard model, in a dynamical mean-field theory.
The $H-T$ phase
diagram  obtained
for an intermediate on-site interaction $U = 2$ is qualitatively
very similar to
that of FeBr$_2$.  Apart
from calculations at $U < 2$, it will be interesting to calculate
off half-filling. This will allow us to investigate the properties
of metallic metamagnets, such as the Uranium-based mixed-systems, for
which a theory in terms of an itinerant electron model is mandatory.

We acknowledge useful correspondence with N. Giordano,
G. Lander, B. L\"{u}thi, A. Ramirez, and W. Wolf, and are grateful to
H. Capellmann, P. van Dongen,  W. Metzner, H. M\"{u}ller-Krumbhaar,
F.~Steglich, G. Stewart
and, in particular, to V. Dohm, W. Kleemann and W. Selke for very helpful
discussions. This work was supported in part by the SFB 341
of the DFG and by a grant from the Office of Naval Research,
ONR N00014--93--1--0495.

\newpage


\section*{Figure Captions}

\begin{enumerate}
\item
Schematic $H-T $ phase diagram for
a) a typical Ising-type metamagnet (TCP: tricritical point),
b) the Ising model (\ref{Gl1}) in
mean-field theory with $R < 3/5$ (CE: critical endpoint, BCE:
bicritical endpoint). Full lines: 1.~order transition,
broken lines: 2.~order
transition; AF: antiferromagnetic phase, P: paramagnetic phase.

\item
QMC-results, including error bars,
for the magnetization $m(H)$ and staggered magnetization
$m_{st}(H)$ as obtained for the $d = \infty$
Hubbard model with easy axis along $H$ at $n = 1$ and $U = 2$
for two sets of temperatures. Curves are guides to the eye only.

\item
$H-T$ phase diagram for
the $d = \infty$ Hubbard model with easy axis along $H$ at $n = 1$
and $U = 2$
as constructed from the QMC-results for $m(H)$ and $m_{st}(H)$;
same notation as in Fig.~1b.

\end{enumerate}


\begin{references}
\bibitem{26} Here $m$ and
$m_{st}$ are defined in terms of the sublattice magnetization
$m_{\gamma} = (2/L) \sum_{i \in \gamma}
\langle \hat{n}_{i \uparrow} -   \hat{n}_{i \downarrow} \rangle$, with
$\gamma = A$ or $B$, and $L$ as the number of lattice sites, by
$m = (m_A + m_B)/2$ and $m_{st} = (m_A- m_B)/2$.~
\bibitem{27} Note that the mixed-phase region in the phase diagram
of Ref.~12 disappears when instead of the applied field $H_a$ the
internal field $H$ is plotted vs.~$T$.
\end{references}
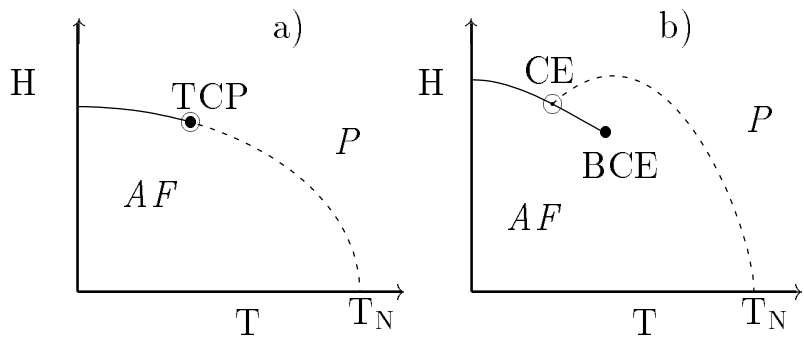
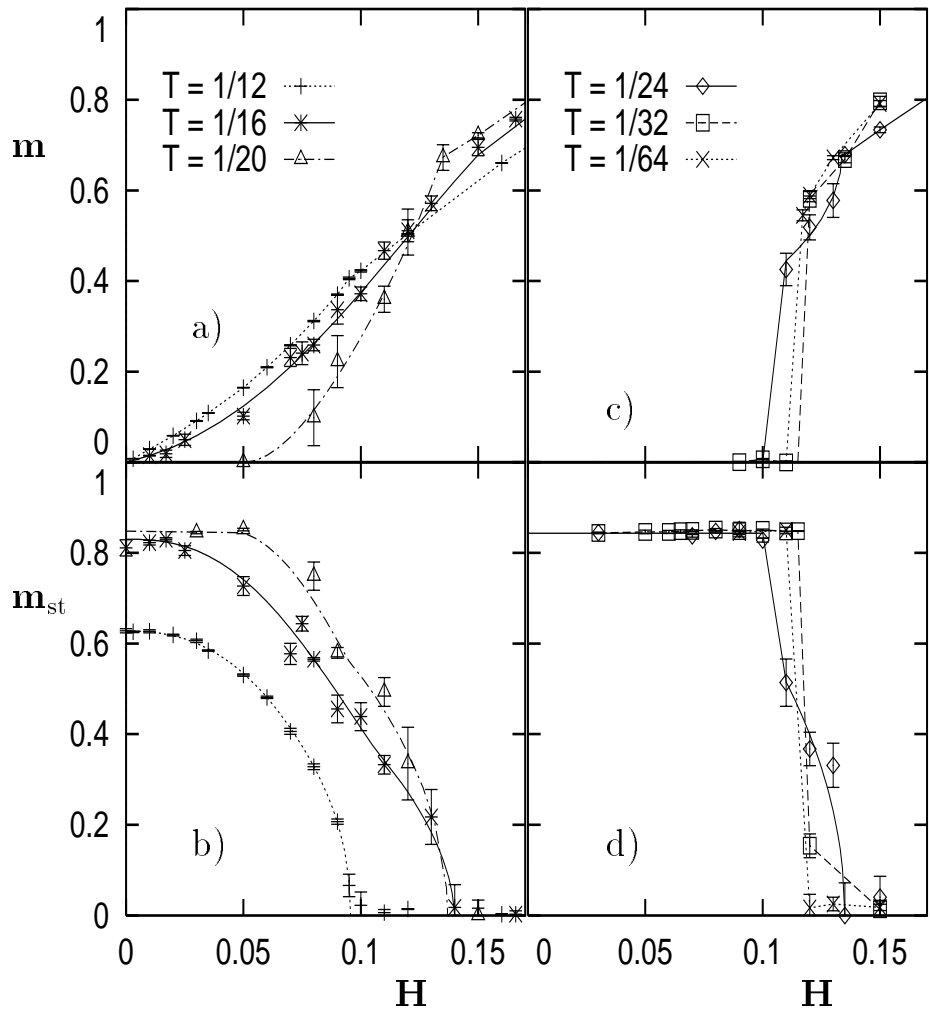
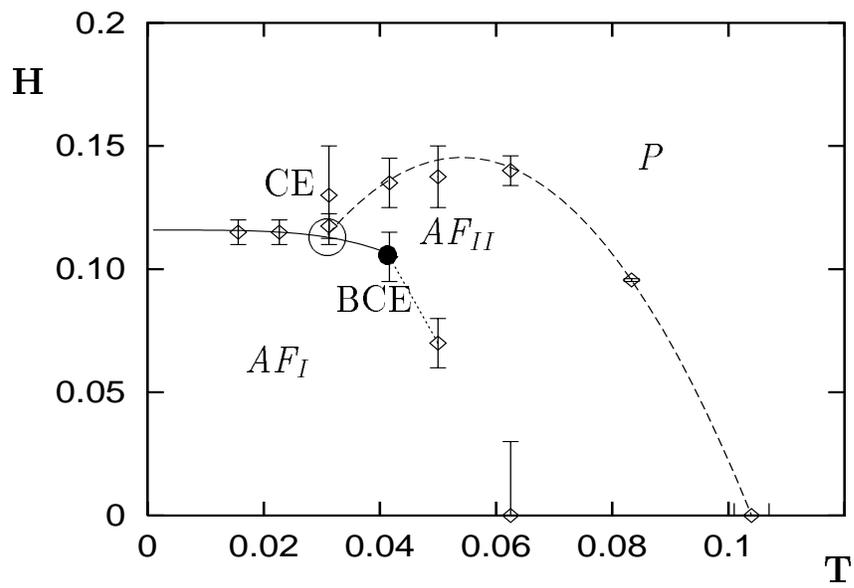
\end{document}